\documentclass[12pt]{article}
\usepackage{epsfig}
\usepackage{lineno}
\usepackage{graphicx}
\usepackage{amsmath}
\usepackage{hhline}
\usepackage{amssymb}
\usepackage{times}
\usepackage{cite}
\usepackage{array}
\usepackage{multirow}
\def\rpv{\mbox{$\rlap{\kern0.25em/}R_p$ }}

\begin{document}

%\begin{titlepage}

%\vspace*{2cm}

%\begin{center}
%\begin{Large}

\title  { Effects of R-parity violation in unpolarized top quark
decay into polarized W-boson } \vspace{3cm}
\author{{Yi Min Nie, Chong Sheng Li\footnote{csli@pku.edu.cn}, Qiang Li, Jian Jun Liu and Jun Zhao  }\\
\small $Department$ $of$ $Physics$, $Peking$ $University$,
$Beijing$ $100871$, $China$}
%\end{Large}
%\end{center}

\vspace{2cm} \maketitle %\vspace{.2in}
\begin{footnotesize}
%\begin{center}\begin{minipage}{5in}
%\baselineskip=0.25in
\begin{abstract} \noindent

%\begin{center} ABSTRACT \end{center}
We calculate one-loop R-parity violating corrections to the top
quark decay into a bottom quark and a polarized W-gauge boson. The
corrections are presented according to the total corrections, the
longitudinal corrections and the transverse corrections,
respectively. We compared our results with the ${\cal
O}(\alpha_s)$ QCD corrections, the ${\cal O}(\alpha)$ electroweak
(EW) and finite width corrections, and also the supersymmetric
(SUSY) corrections with R-parity conservation, respectively.
%We find that the total corrections
%$\delta\Gamma/\Gamma_0$ induced by the R-parity violation
%couplings $\lambda''$ can reach about $4\%$, which are smaller
%than the ${\cal O}(\alpha_s)$ radiative corrections, but are
%comparable with the corresponding results in the minimal
%supersymmetric standard model(MSSM) with R-parity conservation. In
%addition, The longitudinal corrections can reach about $3\%$,
%which are still smaller than the ${\cal O}(\alpha_s)$ QCD
%corrections, but larger than both of the ${\cal O}(\alpha)$
%electroweak(EW) corrections and the supersymmetric(SUSY)
%corrections with R-parity conservation. The magnitudes of the
%corrections to the transverse-minus are about $0.3\%$, which are
%comparable with the SUSY corrections with R-parity conservation,
%but are smaller than both of the ${\cal O}(\alpha_s)$ QCD
%corrections and the ${\cal O}(\alpha)$ EW corrections.
\end{abstract}
%\end{minipage}\end{center}

\vspace{4cm}
%\end{titlepage}
\end{footnotesize}


\newpage

%&&&&&&&&&&&&&&&&&&&&&&&&&&&&&&&&&&&&&&&&&&&&&&&&&&&&&&&&&&&&&&&&&&&&&&&&&&&&&&&&&&&
\section{\bf Introduction}
\noindent The top quark is the heaviest known fermion with a mass
close to the scale of the electroweak(EW) symmetry breaking.
Therefore the study of its properties and the possible deviations
from standard model (SM) predictions can probe varies physics
beyond the SM. The Tevatron run I has yielded relatively small
numbers of top quark events. But the Tevatron run II can provide
copious top quark events, and improve the precision of the top
quark measurements, moreover, the CERN Large Hadron Collider (LHC)
and the Next-generation Linear Collider (NLC) will serve as top
quark factories, and these machines should be very useful to
obtain the information of new physics through analyzing the
properties of the top quark decay, especially the study of the
polarized properties of the decay products will be a good probe of
new physics.

The dominate decay mode of the top quark is $t\rightarrow bW^+$,
where the W boson as a decay product is strongly polarized with
three helicity contents: transverse plus, transverse minus and
longitudinal. The recent measurement of the top quark mass is
$m_t=178.0\pm4.3$ GeV \ \ (world \ \ average)\cite{c1}, which
implies
\begin{eqnarray} \Gamma=1.86 GeV \ \ (including \ \ {\cal O}(\alpha_s) \
\ \ QCD \ \ \ corrections),
\end{eqnarray}
and the helicity of the W boson measured by CDF
Collaborations\cite{c2} before the publication of \cite{c1} is
\begin{eqnarray}
\Gamma_L/\Gamma=0.91\pm0.37(stat)\pm 0.13(syst),
\end{eqnarray}
and
\begin{eqnarray}
\Gamma_+/\Gamma=0.11\pm 0.15 \ \ \ \ \ \ (assuming\ \ \ \ \
\Gamma_L/\Gamma=0.7),
\end{eqnarray}
where $\Gamma$ is the total width of $t\rightarrow bW^+$ ,
$\Gamma_L$ and $\Gamma_+$ are the longitudinal and the
transverse-plus widths, respectively. Since the size of the
longitudinal contribution encodes the physics of the spontaneous
breaking of the electroweak symmetry, and the transverse-plus
contribution vanishes at the Born-level, any sizable deviation
from them may indicate obvious quantum effects or a non-SM (V$+$A)
coupling in the weak $t\rightarrow b$ current transition.
Therefore, the investigation of radiative corrections to the top
decay into a polarized W boson may provide additional information
about the new physics, and there has been a great of interests
from theorists. So far theoretical predictions of these
observables reported in the literatures are summarized as follows:
the QCD corrections to the top width are
rather large in general. One-loop QCD corrections to tree-level width are  $ -8.54\% $~%
\cite{c3,c4,c5,c6,c7,c8},  one-loop electroweak corrections are $ +1.54\% $ ~%
\cite{c3,c9,c10} and two-loop QCD corrections are approximate $
-2.05\% $ ~\cite{c11} ($ -2.16\% $ \hspace{-2mm}\ \ \ \cite{c12}\
), and the one-loop SUSY-QCD and SUSY-EW corrections are below
$1\%$ in magnitude for most of the parameter space\cite{c15}. As
for the partial longitudinal and transverse rates, for the top
quark mass $ m_t=175GeV $, the tree-level results are 0.703 for
$\Gamma_L/\Gamma$, 0.297 for $\Gamma_-/\Gamma$ and ${\cal
O}(10^{-4})$ for $\Gamma_+/\Gamma$, respectively. In the SM, the
${\cal O}(\alpha_s)$ QCD corrections decrease the rate
$\Gamma_L/\Gamma$ by $1.06\%$, increase the rate $\Gamma_-/\Gamma$
by $2.17\%$ and the rate $\Gamma_+/\Gamma$ reach mere
$0.10\%$\cite{c13}. The electroweak and finite width corrections
increase the rate $\Gamma_L/\Gamma$ by $1.32\%$, increase the rate
$\Gamma_-/\Gamma$ by $2.06\%$ and the rate $\Gamma_+/\Gamma$ are
only ${\cal O}(0.1\%)$\cite{c14}. Beyond the SM, one-loop SUSY-QCD
and SUSY-EW corrections to $\Gamma_-/\Gamma$ and $\Gamma_L/\Gamma$
are less than $1\%$ in magnitude and tend to have opposite
signs\cite{c15}. However, the R-parity violating SUSY
contributions to $t\rightarrow bW^+$ have not been calculated so
far. In this paper we will investigate the R-parity violating SUSY
contributions to $t\rightarrow bW^+$, including the total
corrections, the longitudinal corrections and the transverse
corrections, respectively.

The most general superpotential of the minimal supersymmetric
standard model (MSSM) consistent with the $SU(3)\times SU(2)\times
U(1)$ symmetry and supersymmetry contains R-violating
interactions, which are given by\cite{c16}
\begin{eqnarray}
{\cal W}_{\not
R}=\frac{1}{2}\lambda_{ijk}L_iL_jE_k^c+\lambda'_{ijk}\delta^{\alpha\beta}L_iQ_{j\alpha}D^c_{k\beta}
+\frac{1}{2}\lambda^{''}_{ijk}\varepsilon^{\alpha\beta\gamma}U^c_{i\alpha}D^c_{j\beta}D^c_{k\gamma}+\mu_iL_iH_2.
\end{eqnarray}
\noindent Here $L_i (Q_i)$ and $E_i (U_i, D_i)$ are, respectively,
the left-handed lepton (quark) doublet and right-handed lepton
(quark) singlet chiral superfields, and  $H_{1,2}$ are the Higgs
chiral superfields. The indices $i, j, k$ denote generations and
$\alpha$ $\beta$ and $\gamma$ are the color indices, and the
superscript $c$ denotes charge conjugation. The $\lambda$ and
$\lambda'$ are the coupling constants of  L(lepton
number)-violating interactions and $\lambda^{''}$ those of
B(baryon number)-violating interactions. The non-observation (so
far) of the proton decay imposes very strong constraints on the
product of L-violating and B-violating couplings. It is thus
conventionally assumed in the phenomenological studies that only
one type of these interactions (either L- or B-violating) exists.
Some constraints on these R-parity violating couplings have been
obtained from various analysis of their phenomenological
implications based on experiments. It is notable that the bounds
on the couplings involving top quark are generally quite weak. As
the large number of top quarks will be produced at the future
colliders, these couplings may either manifest themselves, or
stronger constraints on them can be established. For the purpose
of our work, we will focus on the quantum effects of R-parity
violating couplings involved $\lambda^{''}_{ijk}$ and
$\lambda^{'}_{ijk}$ in the top quark decay into polarized $W^+$
boson, respectively.

% It is important that the R-parity violation is much
%less constrained in general for the $3$-rd fermion-sfermion
%generation compared to the first two generations.

This paper is organized as follows. In the Sec.2 we present some
formulas for our calculations, by which we can calculate the
helicity amplitude of the top quark decays. In the Sec.3 we take a
careful study on R-parity violating effects using the helicity
method. In the Sec.4 we make a numerical analysis with drawing
some conclusions for our calculations. The explicit expressions of
the helicity amplitudes induced by the R-parity violating
couplings are given in the Appendix.

%&&&&&&&&&&&&&&&&&&&&&&&&&&&&&&&&&&&&&&&&&&&&&&&&&&&&&&&&&&&&&&&&&&&&&&&&&&&&&&&&&&&&&&&&&&&&&&&&&&&&&&&&&&&
\section{\bf Formalism}

In order to make our paper self-constrained, we start with a brief
description of the helicity amplitude method for performing the
following calculations. The method breaks down the algebra of
four-dimensional Dirac spinors and matrices into equivalent
two-dimensional ones. In what follows we introduce the Weyl
representation of Dirac spinors and matrices. In spherical
coordinates the four-momemta can be written as
\begin{eqnarray}
p^\mu=(E,|\vec{p} |\sin\theta\cos\varphi,|\vec{p}
|\sin\theta\sin\varphi,|\vec{p} |\cos\theta),
\end{eqnarray}
with $E^2-| \vec{p}^2 |=m^2$. The left-hand (L), right-hand (R)
and longitudinal (0) polarization vectors for a spin-1 field are
defined as \cite{Haber}
\begin{eqnarray}
\varepsilon_{(L)}^\mu&=&\frac{e^{-i\phi}}{\sqrt{2}}(0,i\sin\phi+\cos\phi\cos\theta,-i\cos\phi+\sin\phi\cos\theta,-\sin\theta),\nonumber\\
\varepsilon_{(R)}^\mu&=&\frac{e^{i\phi}}{\sqrt{2}}(0,i\sin\phi-\cos\phi\cos\theta,-i\cos\phi-\sin\phi\cos\theta,\sin\theta),\\
\varepsilon_{(0)}^\mu&=&\frac{1}{m}(|\vec{p}|,E\sin\theta\cos\phi,E\sin\theta\sin\phi,E\cos\theta).\nonumber
\end{eqnarray}
The above equations satisfy the identities
$\varepsilon_{(R)}^\mu=-\varepsilon_{(L)}^{\mu*},\varepsilon_{(0)}^\mu=\varepsilon_{(0)}^{\mu*},p_\mu\varepsilon_{(h)}^\mu=0$
and $\varepsilon^{(h)}_\mu \varepsilon^{\mu
*}_{(h^\prime)}=-\delta_{hh^\prime}$ for $h,h^\prime=R,L$ or $0$.
In the Weyl basis Dirac spinors have the following four component
forms
\begin{eqnarray}
\psi=\left( \begin{array}{cc} \psi_+\\\psi_-\end{array} \right).
\end{eqnarray}
For fermions (with helicity $\lambda=\pm1$), \begin{eqnarray}
\psi_{\pm}=\bigg\{\begin{array}{cc}
u_{\pm}^{(\lambda=1)}=\omega_{\pm}\chi_{1/2}\\u_{\pm}^{(\lambda=-1)}=\omega_{\mp}\chi_{-1/2}\end{array},
\end{eqnarray}
and for anti-fermions (with helicity $\lambda=\pm1$),
\begin{eqnarray}
\psi_{\pm}=\bigg\{\begin{array}{cc}
v_{\pm}^{(\lambda=1)}=\pm\omega_{\mp}\chi_{-1/2}\\v_{\pm}^{(\lambda=-1)}=\mp\omega_{\pm}\chi_{1/2}\end{array},\end{eqnarray}
with $\omega_{\pm}=\sqrt{E\pm|\vec{p}|}$. The $\chi_{\lambda/2}$'s
are eigenvectors of the helicity operator
\begin{eqnarray}h=\vec{p}\cdot\vec{\sigma},  \    \   \     \hat{p}=\vec{p}/|\vec{p}|,\end{eqnarray}
where $\sigma_{j=1,2,3}$ are the Pauli matrices. We use the top
rest frame, and one can write down these eigenvectors with
eigenvalue $\lambda$ (for simplicity we take $\phi=0$) as
\begin{eqnarray}\chi_{1/2}=\left( \begin{array}{cc}
\cos\frac{\theta}{2}\\\sin\frac{\theta}{2}
\end{array}\right), \    \     \
\chi_{-1/2}=\left( \begin{array}{cc}
-\sin\frac{\theta}{2}\\\cos\frac{\theta}{2}
\end{array}\right),\end{eqnarray}
where $\lambda=\pm1$ are for "spin-up" and "spin-down" ,
respectively. In the Weyl basis \mbox{$\rlap{\kern0.1em/}p$ }
takes the form
\begin{eqnarray}\not p=p_\mu\gamma^\mu\equiv\left( \begin{array}{cc} 0&\not p_+\\\not p_-&0
\end{array}\right)\equiv p_\mu\left( \begin{array}{cc} 0&\gamma_+^\mu\\\gamma_-^\mu&0
\end{array}\right),\end{eqnarray}
where
\begin{eqnarray}\gamma_\pm^\mu=(1,\pm\vec{\sigma}).\end{eqnarray}
According to the above discussions, in the top rest frame we can
write down the following expressions:
\begin{eqnarray}
p^\mu_t&=&(m_t,0,0,0),\nonumber\\
p_W^\mu&=&(E_W,-|\vec{p}_W|\sin\theta,0,-|\vec{p}_W|\cos\theta),\nonumber\\
\varepsilon^\mu_{(+)}&=&\frac{1}{\sqrt{2}}(0,\cos\theta,-i,-\sin\theta),\\
\varepsilon^\mu_{(0)}&=&\frac{1}{M_W}(|\vec{p}_W|,-E_W\sin\theta,0,-E_W\cos\theta),\nonumber\\
\varepsilon^\mu_{(-)}&=&-(\varepsilon^\mu_{(+)})^*,\nonumber
\end{eqnarray}
where $p^\mu_t$ and $p^\mu_W$ are the four-dimensional momenta of
the top quark and W boson,
$\varepsilon^\mu_{(+)},\varepsilon^\mu_{(-)}$ and
$\varepsilon^\mu_{(0)}$ denote the left-hand (L), right-hand (R)
and longitudinal (0) polarization vectors of W boson,
respectively. Moreover, $E_W$ is the energy of the W boson and
$\vec{p}_b$ is the momentum of the bottom quark. They can be
expressed as
\begin{eqnarray}
E_W&=&\frac{m_t^2+m_b^2-M_W^2}{2m_t},\\
|\vec{p}_b|&=&\frac{1}{2m_t}\sqrt{[(m_t+m_b)^2-M_W^2][(m_t-m_b)^2-M_W^2]}.
\end{eqnarray}
%And the energy of bottom quark is
%\begin{eqnarray}E_b&=&\frac{m_t^2+M_W^2-m_b^2}{2m_t}.\nonumber\end{eqnarray}
\section{\bf Calculations}
\subsection{\bf Tree level results} The tree level amplitude, as shown in
Fig.1(a), is given by
\begin{eqnarray}
A_0=\bar{u}(p_b)\gamma^\mu[\frac{-ig}{\sqrt{2}}(\frac{1-\gamma_5}{2})]u(p_t)\varepsilon_\mu(p_W),
\end{eqnarray}
where $\varepsilon_\mu$ is the polarization vector of W boson.
According to the previous discussions in Section II, the
corresponding helicity amplitude can be written as
\begin{eqnarray}
A_0(\lambda_t,\lambda_b,\lambda_W)=\frac{-ig}{\sqrt{2}}u^\dag_-(p_b,\lambda_b)\gamma^\mu_+u_-(p_t,\lambda_t)\varepsilon_\mu(p_W,\lambda_W).\nonumber
\end{eqnarray}
According to Eq.(12), the spinor of the bottom quark is
\begin{eqnarray}
u_-(\lambda_b=-1)&=&\omega_+(b)\chi_{-\frac{1}{2}}(b)=\sqrt{E_b+|\vec{p}_b|}\left(
\begin{array}{cc} -\sin\frac{\theta}{2}\\\cos\frac{\theta}{2}
\end{array}\right),\\
u_-(\lambda_b=+1)&=&\omega_-(b)\chi_{\frac{1}{2}}(b)=\sqrt{E_b-|\vec{p}_b|}\left(
\begin{array}{cc} \cos\frac{\theta}{2}\\\sin\frac{\theta}{2}
\end{array}\right),
\end{eqnarray}
where
\begin{eqnarray}E_b&=&\frac{m_t^2+M_W^2-m_b^2}{2m_t}.\nonumber\end{eqnarray}
And the spinors of the spin-down(up) top quark are shown as below,
respectively,
\begin{eqnarray}
u_-(\lambda_t=-1)&=&\omega_+(t)\chi_{-\frac{1}{2}}(t)=\sqrt{E_t}\left(
\begin{array}{cc} 0\\1
\end{array}\right),\\
u_-(\lambda_t=+1)&=&\omega_-(t)\chi_{\frac{1}{2}}(t)=\sqrt{E_t}\left(
\begin{array}{cc} 1\\0
\end{array}\right).
\end{eqnarray}
Thus the explicit expressions of the helicity amplitudes
$A(\lambda_t,\lambda_b,\lambda_W)$ are given by
\begin{eqnarray}
& &A_0(+,-,0)=\frac{ig}{\sqrt{2}M_W}\sqrt{(E_b+|\vec{p}_b|)E_t}(E_W+|\vec{p}_b|)\sin\frac{\theta}{2},\nonumber\\
& &A_0(+,-,-1)=-ig\sqrt{(E_b+|\vec{p}_b|)E_t}\cos\frac{\theta}{2},\nonumber\\
& &A_0(+,-,+1)=0,\nonumber\\
& &A_0(-,-,0)=\frac{-ig}{\sqrt{2}M_W}\sqrt{(E_b+|\vec{p}_b|)E_t}(E_W+|\vec{p}_b|)\cos\frac{\theta}{2},\nonumber\\
& &A_0(-,-,-1)=-ig\sqrt{(E_b+|\vec{p}_b|)E_t}\sin\frac{\theta}{2},\nonumber\\
& &A_0(-,-,+1)=0.\nonumber\\& &A_0(+,+,0)=\frac{ig}{\sqrt{2}M_W}(E_W-|\vec{p}_b|)\sqrt{(E_b-|\vec{p}_b|)E_t}\cos\frac{\theta}{2},\nonumber\\
& &A_0(+,+,-1)=0,\nonumber\\
& &A_0(+,+,+1)=-ig\sqrt{(E_b-|\vec{p}_b|)E_t}\sin\frac{\theta}{2},\nonumber\\
& &A_0(-,+,0)=\frac{-ig}{\sqrt{2}M_W}(E_W-|\vec{p}_b|)\sqrt{(E_b-|\vec{p}_b|)E_t}\sin\frac{\theta}{2},\nonumber\\
& &A_0(-,+,-1)=0,\nonumber\\
&
&A_0(-,+,+1)=-ig\sqrt{(E_b-|\vec{p}_b|)E_t}\cos\frac{\theta}{2}.\nonumber
\end{eqnarray}
In this paper we only consider the longitudinal width $\Gamma_L$
or the transverse width $\Gamma_T$ induced by the polarization of
the $W^+$, so we should sum the helicities of the top quark and
the bottom quark. Using the previous results, the tree-level ratio
of $\Gamma_L$ and $\Gamma_T$ is
\begin{eqnarray}
\frac{\Gamma_L}{\Gamma_{T}}=\frac{\sum_{\lambda_t,\lambda_b}|A_0|^2(\lambda_t,\lambda_b,\lambda_W=0)}{\sum_{\lambda_t,\lambda_b}|A_0|^2(\lambda_t,\lambda_b,\lambda_W=\pm1)}=\frac{M_W^2+2|\vec{p}_b|^2+2\frac{E_W}{E_b}|\vec{p}_b|^2}{2M_W^2}.
\end{eqnarray}
If the mass of the bottom quark is neglected, Eq.(21) is
simplified as
\begin{eqnarray}\frac{\Gamma_L}{\Gamma_T}=\frac{m_t^2}{2M_W^2}.\end{eqnarray}
\subsection{SUSY R-Parity violation corrections}
The SUSY R-parity violation corrections arise from the Feynman
diagrams shown in Fig1(b)-($c'$) and Fig (2)-(4), which consist of
the vertex and self-energy diagrams. The Lagrangians involved the
$\lambda^{''}$ and $\lambda'$ couplings are given by
\begin{eqnarray}
{\cal
L}_{UDD}&=&-\lambda^{''}_{ijk}\varepsilon_{\alpha\beta\gamma}[(\tilde{d}_R^{kr})^*\bar{u}_R^{i\alpha}(d_R^{j\beta})^c
+\frac{1}{2}(\tilde{u}_R^{i\alpha})^*\bar{d}_R^{j\beta}(d_R^{k\gamma})^c]+h.c.,\nonumber\\
{\cal
L}_{LQD}&=&-\lambda'_{ijk}(\tilde{d}^*_{kR}\bar{v}^c_iP_Ld_j-\tilde{d}^*_{kR}\bar{l}^c_iP_Lu_j+
\tilde{d}_{jL}\bar{d}^kP_Lv_i\\&
&-\tilde{u}_{jL}\bar{d}^kP_Ll_i+\tilde{v}_{i}\bar{d}^kP_Ld_j-
\tilde{l}_{iL}\bar{d}^k_iP_Lu_j)+h.c. .\nonumber
\end{eqnarray}
Note that $\lambda^{''}_{ijk}=-\lambda^{''}_{ikj}$. When we
consider the one-loop SUSY R-parity violation corrections, the
renormalized amplitude can be written as
\begin{eqnarray}
& &A^{ren}=A^0+\delta A,\\
& &\delta A \equiv A^v+A^c,
\end{eqnarray}
where $A^v$ and $A^c$ are the vertex corrections and the
conterterms, respectively.
\begin{itemize}
\item {\bf The corrections from the $\lambda''$ couplings}\\
Calculating the diagrams in Fig.1 $(b)$ and $(c)$, we can get the
explicit expressions of the vertex corrections as following:
\begin{eqnarray}
A_1^v&=&\sum_{j,k=1}^{3}\sum_{m=1}^{2}\bigg\{\frac{igX}{8\sqrt{2}\pi^2}\bar{u}(p_b)\gamma^{\mu}
P_{R}u(p_t)\varepsilon_\mu(p_W)
C_0(m_t^2,m_b^2,M_W^2,m^2_{d_j},m^2_{u_j},m^2_{\tilde{d}_{k,m}})\bigg\},\nonumber\\\\
A_2^v&=&\sum_{j,k=1}^{3}\sum_{m,n=1}^{2}\bigg\{\frac{igY}{16\sqrt{2}\pi^2}[
m_t\bar{u}(p_b)P_Lu(p_t)C_\mu\varepsilon^\mu(p_W)\nonumber\\
& &+2\bar{u}(p_b)P_L\gamma^\mu C_{\mu\nu}\varepsilon^\nu(p_W)
u(p_t)](m_t^2,m_b^2,M_W^2,m^2_{d_j},m^2_{\tilde{u}_{k,n}},m^2_{\tilde{d}_{k,m}})\bigg\},
\end{eqnarray}
with
\begin{eqnarray}
X&=&\lambda^{''}_{3jk}\lambda^{''}_{j3k}| R^{\tilde{d_k}}_{m2}|^2
m_{u_j}m_{d_j},\nonumber\\
Y&=&-\lambda^{''}_{3jk}\lambda^{''}_{kj3}R^{\tilde{u_k}}_{n1}R^{\tilde{u_k}}_{n2}
R^{\tilde{d_k}}_{m1}R^{\tilde{d_k}}_{m2},\nonumber
\end{eqnarray}
where $C_0$, $C_{\mu}$ and $C_{\mu\nu}$ are the three-point
integrals, which are defined similar to\cite{c17} except that we
take internal masses squared as arguments\cite{c18}.
$m_{\tilde{u}_k(\tilde{d}_k)_{1,2}}(k=1,2,3)$ are the squark
masses, and $R^{\tilde{u}_k(\tilde{d}_k)}$ are $2\times 2$ matrix,
which are defined to transform the squark current eigenstates to
the mass eigenstates.

The corresponding helicity amplitudes are
\begin{eqnarray}
A^v_1(\lambda_t,\lambda_b,\lambda_W)&=&\sum_{j,k=1}^{3}\sum_{m=1}^{2}\bigg\{\frac{igX}{8\sqrt{2}\pi^2}
u^\dagger_+(p_b,\lambda_b)\gamma^{\mu}_-u_+(p_t,\lambda_t)\varepsilon_\mu(p_W,\lambda_W)\nonumber\\&
&
\times C_0(m_t^2,m_b^2,M_W^2,m^2_{d_j},m^2_{u_j},m^2_{\tilde{d}_{k,m}})\bigg\},\\
A_2^v(\lambda_t,\lambda_b,\lambda_W)&=&\sum_{j,k=1}^{3}\sum_{m,n=1}^{2}\bigg\{\frac{igY}{16\sqrt{2}\pi^2}[
m_tu_-^\dag(p_b,\lambda_b)u_-(p_t,\lambda_t)C_\mu\varepsilon^\mu(p_W,\lambda_W)\nonumber\\
& &+2u^\dagger_+(p_b,\lambda_b)\gamma^{\mu}_-u_+(p_t,\lambda_t)
C_{\mu\nu}\varepsilon^\nu(p_W,\lambda_W)](m_t^2,m_b^2,M_W^2,m^2_{d_j},m^2_{\tilde{u}_{k,n}},m^2_{\tilde{d}_{k,m}})\bigg\},\nonumber\\
\end{eqnarray}
the explicit expressions of which are shown in the Appendix.

The counterterm $A^c$ can be expressed as
\begin{eqnarray}
A^c=(\frac{1}{2}\delta Z^t_{L1}+\frac{1}{2}\delta Z^b_{L1})A^0,
\end{eqnarray}
where $\delta Z^t_{L1}, \delta Z^b_{L1}$ are the renormalization
constants, which are fixed by the on-shell renormalization
scheme\cite{c19} and can be got from the calculations of the
self-energy diagrams in Fig.2:
\begin{eqnarray}
\delta Z^t_{L1}&=&-\sum_{j,k=1}^{3}\sum_{m=1}^{2}\bigg\{\frac{im_t^2}{8\pi^2}|\lambda^{''}_{3jk}|^2|R^{\tilde{d_k}}_{m2}|^2B'_1(p_t^2,m^2_{d_j},m^2_{\tilde{d}_{k,m}})\bigg\},\\
%\delta Z^t_{R1}&=&\Sigma^t_R(m_t^2)+m_t^2[2\Sigma_s^{t'}(m_t^2)+\Sigma_L^{t'}(m_t^2)+\Sigma_R^{t'}(m_t^2)]\nonumber\\
%&=&-\frac{i}{8\pi^2}|\lambda^{''}_{3jk}|^2|R^{\tilde{d_k}}_{m2}|^2[B_1(p_t^2,m^2_{d_j},m^2_{\tilde{d}_{k,m}})+m_t^2B'_1(p_t^2,m^2_{d_j},m^2_{\tilde{d}_{k,m}})]
\delta
Z^b_{L1}&=&-\sum_{i,k=1}^{3}\sum_{m=1}^{2}\bigg\{\frac{im_b^2}{8\pi^2}|\lambda^{''}_{i3k}|^2|R^{\tilde{d_k}}_{m2}|^2B'_1(m_b^2,m^2_{u_i},m^2_{\tilde{d}_{k,m}})\nonumber\\
& &+\frac{im_b^2}{32\pi^2}
|\lambda^{''}_{i3k}|^2|R^{\tilde{u_i}}_{m2}|^2B'_1(m_b^2,m^2_{\tilde{u}_{i,m}},m^2_{d_k})\bigg\},
%\delta
%Z^b_{R1}&=&-\frac{i}{8\pi^2}|\lambda^{''}_{i3k}|^2|R^{\tilde{d_k}}_{m2}|^2[B'_1(m_b^2,m^2_{u_i},m^2_{\tilde{d_k}})+m_b^2B'_1(m_b^2,m^2_{u_i},m^2_{\tilde{d_k}})]\nonumber\\&-&\frac{i}{32\pi^2}
%|\lambda^{''}_{i3k}|^2|R^{\tilde{u_i}}_{m2}|^2[B'_1(m_b^2,m^2_{u_i},m^2_{\tilde{d_k}})+m_b^2B'_1(m_b^2,m^2_{\tilde{u_i}},m^2_{d_k})]
\end{eqnarray}
where $B'_1=\partial B_1/\partial p^2$, and $B_1$ is the two-point
integral\cite{c18}.

The one-loop SUSY R-parity violation corrections to the decay
width are given by
\begin{eqnarray}
\delta \Gamma_{\lambda_W} =2{\rm Re}
\bigg\{\frac{1}{2}{\sum_{\lambda_t,\lambda_b}}{A^{(0)\dag}(\lambda_t,\lambda_b,\lambda_W)
\delta A(\lambda_t,\lambda_b,\lambda_W)\bigg\}}\cdot PS\ \
\nonumber,
\end{eqnarray}
where $\lambda_W=(L,+,-)$, and $PS$ is given by
\begin{eqnarray}
PS=\frac{1}{16\pi
m_t^3}\sqrt{[m_t^2-(M_W+m_b)^2][m_t^2-(M_W-m_b)^2]}.
\end{eqnarray} The ratios
$\delta\hat{\Gamma}_{\lambda_W}=\delta\Gamma_{\lambda_W}/\Gamma_0
({\lambda_W}=L,+,-)$ are then given by
\begin{eqnarray}
\delta\hat{\Gamma}_L&=&\sum_{j,k=1}^{3}\sum_{m,n=1}^{2}\bigg\{\frac{M_W^2}{m_t[(3m_W^2+2|\vec{p}_b|^2)E_b+2E_W|\vec{p}_b|^2]}
\times\nonumber\\&
&\big\{\frac{m_tm_b}{4\pi^2}[-XC_0(m_t^2,m_b^2,M_W^2,m^2_{d_j},m^2_{u_j},m^2_{\tilde{d}_{k,m}})\nonumber\\
& &+\frac{m_t^2}{M_W^2}|\vec{p}_b|^2Y(C_{12}+C_{22})
(m_t^2,m_b^2,M_W^2,m^2_{d_j},m^2_{\tilde{u}_{k,n}},m^2_{\tilde{d}_{k,m}})]+\nonumber\\&
&\frac{m_t[(3M_W^2+2|\vec{p}_b|^2)E_b+2E_W|\vec{p}_b|^2]}{M_W^2}(\delta
Z^t_{L1}+\delta Z^b_{L1})\big\}\bigg\}\\
\delta\hat{\Gamma}_-&=&\sum_{j,k=1}^{3}\sum_{m,n=1}^{2}\bigg\{\frac{M_W^2}{[(3m_W^2+2|\vec{p}_b|^2)E_b+2E_W|\vec{p}_b|^2]}\times \nonumber\\
& &\big\{(E_b+|\vec{p}_b|)(\delta Z^t_{L1}+\delta
Z^b_{L1})+\frac{m_b}{4\pi^2}[XC_0(m_t^2,m_b^2,M_W^2,m^2_{d_j},m^2_{u_j},m^2_{\tilde{d}_{k,m}})\nonumber\\& &+YC_{24}(m_t^2,m_b^2,M_W^2,m^2_{d_j},m^2_{\tilde{u}_{k,n}},m^2_{\tilde{d}_{k,m}})]\big\}\bigg\},\\
\delta\hat{\Gamma}_+&=&\sum_{j,k=1}^{3}\sum_{m,n=1}^{2}\bigg\{\frac{M_W^2}{[(3m_W^2+2|\vec{p}_b|^2)E_b+2E_W|\vec{p}_b|^2]}\times\nonumber\\
& &\big\{(E_b-|\vec{p}_b|)(\delta Z^t_{L1}+\delta
Z^b_{L1})+\frac{m_b}{4\pi^2}[XC_0(m_t^2,m_b^2,M_W^2,m^2_{d_j},m^2_{u_j},m^2_{\tilde{d}_{k,m}})\nonumber\\&
&+YC_{24}(m_t^2,m_b^2,M_W^2,m^2_{d_j},m^2_{\tilde{u}_{k,n}},m^2_{\tilde{d}_{k,m}})]\big\}\bigg\}.
\end{eqnarray}
\item {\bf The corrections from the $\lambda'$ couplings} \\
In the case of $\lambda'$, the results can be got in the same way,
and we show the corrections as following:
\begin{eqnarray}
\delta\Gamma_L&=&\sum_{i,k=1}^{3}\sum_{m=1}^{2}\bigg\{\frac{(E_b+|\vec{p}_b|)^2(E_b+|\vec{p}_b|)E_t}{2m_W^2}(\delta
Z^t_{L2}+\delta
Z^b_{L2})\nonumber\\&+&\frac{(E_b+|\vec{p}_b|)|\vec{p}_b|m_bm_t^2}{8\pi^2M_W}|\lambda'_{i3k}|^2
|R^{\tilde{l}_i}_{m1}|^2C_{12}(m_t^2,m_b^2,M_W^2,m^2_{\tilde{d}_{k,m}},0,m^2_{l_{i}})\nonumber\\&-&\frac{(E_b+|\vec{p}_b|)(E_W+|\vec{p}_b|)m_t}{16\pi^2M_W}|\lambda'_{i3k}|^2
|R^{\tilde{d}_k}_{m2}|^2[(E_W+|\vec{p}_b|)|\vec{p}_b|C_{11}
+\frac{m_tE_W(E_W+|\vec{p}_b|)}{M_W}C_{12}\nonumber\\&-&M_W(E_W+|\vec{p}_b|)C_{12}-\frac{(1-Z)(E_W+|\vec{p}_b|)}{M_W}C_{24}
+\frac{|\vec{p}_b|m_t(E_W+|\vec{p}_b|)}{M_W}(C_{22}-C_{23})\nonumber\\&-&\frac{|\vec{p}_b|m_t^2}{M_W}C_{22}](m_t^2,m_b^2,M_W^2,m^2_{\tilde{d}_{k,m}},0,m^2_{l_i})\bigg\},\\
\delta\Gamma_-&=&\sum_{i,k=1}^{3}\sum_{m=1}^{2}\bigg\{\frac{(E_b+|\vec{p}_b|)E_t}{2}(\delta Z^t_{L2}+\delta Z^b_{L2})+\frac{(E_b+|\vec{p}_b|)E_t}{8\pi^2}|\lambda'_{i3k}|^2|R^{\tilde{d}_k}_{m2}|^2[M_W^2C_{11}\nonumber\\
&+&2(1-Z)+(m_tE_W-M_W^2)C_{12}](m_t^2,m_b^2,M_W^2,m^2_{\tilde{d}_{k,m}},0,m^2_{l_i})\bigg\},
\end{eqnarray}
\begin{eqnarray}
\delta\Gamma_+&=&\sum_{i,k=1}^{3}\sum_{m=1}^{2}\bigg\{\frac{(E_b-|\vec{p}_b|)E_t}{2}(\delta Z^t_{L2}+\delta Z^b_{L2})+\frac{(E_b+|\vec{p}_b|)E_t}{8\pi^2}|\lambda'_{i3k}|^2|R^{\tilde{d}_k}_{m2}|^2\times\nonumber\\
&
&[M_W^2C_{11}+2(1-Z)+(m_tE_W-M_W^2)C_{12}](m_t^2,m_b^2,M_W^2,m^2_{\tilde{d}_{k,m}},0,m^2_{l_i})\bigg\}\nonumber\\
\end{eqnarray}
with
\begin{eqnarray}
Z=\big[4C_{24}+M_W^2C_{21}+m_b^2C_{22}+2(m_tE_W-M_W^2)C_{23}\big](m_t^2,m_b^2,M_W^2,m^2_{\tilde{d}_{k,m}},0,m^2_{l_i}),\nonumber
\end{eqnarray}
where $\delta Z^t_{L2}, \delta Z^b_{L2}$ are the renormalization
constants, which can be got from the calculations of the
self-energy diagrams in Fig.3-4:
\begin{eqnarray}
\delta
Z^t_{L2}&=&-\sum_{i,k=1}^{3}\sum_{m=1}^{2}\bigg\{\frac{m_t^2}{16\pi^2}|\lambda'_{i3k}|^2\times\nonumber\\
&
&\big[|R^{\tilde{l}_{i}}_{m1}|^2B'_1(m^2_t,m^2_{\tilde{l}_{i,m}},m^2_{d_k})+
|R^{\tilde{d}_k}_{m2}|^2B'_1(m^2_t,m^2_{\tilde{d}_{k,m}},m^2_{l_i})\big]\bigg\},\\
\delta
Z^b_{L2}&=&-\sum_{i,j,k=1}^{3}\sum_{m=1}^{2}\bigg\{\frac{m_t^2}{16\pi^2}|\lambda'_{ij3}|^2\times\big[|R^{\tilde{d}_{j}}_{m1}|^2B'_1(m^2_b,0,m^2_{\tilde{d}_{j,m}})
\nonumber\\
& &
+|R^{\tilde{u}_{j}}_{m1}|^2B'_1(m^2_b,m^2_{l_i},m^2_{\tilde{u}_{j,m}})+|R^{\tilde{l}_i}_{m1}|^2B'_1(m^2_b,m^2_{\tilde{l}_{i,m}},m^2_{u_j})\big]
\nonumber\\& &
+|\lambda'_{i3k}|^2\big[B'_1(m^2_b,0,m^2_{d_k})+|R^{\tilde{l}_{i}}_{m1}|^2B'_1(m^2_b,0,m^2_{\tilde{d}_{k,m}})\big]\bigg\}.
\end{eqnarray}
\end{itemize}
\section{\bf Numerical Results and Conclusions}
We now present some numerical results for the R-parity violation
effects in the top quark decay into polarized W boson. In our
numerical calculations the SM parameters were taken to be
$\alpha_{ew}=1/128.8$, $m_W=80.419$GeV, $m_t=178.0$GeV ,
$m_Z=91.1882$GeV and $m_b(m_b)=4.25$GeV\cite{c20}.

The relevant SUSY parameters are determined as following:

(i) For the parameters $m^2_{\tilde{Q},\tilde{U},\tilde{D}}$ and
$A_{t,b}$ in squark mass matrices
\begin{eqnarray}
M^2_{\tilde{q}} =\left(\begin{array}{cc} M_{LL}^2 & m_q M_{LR}\\
m_q M_{RL} & M_{RR}^2 \end{array} \right)
\end{eqnarray}
with
\begin{eqnarray}
&&M_{LL}^2 =m_{\tilde{Q}}^2 +m_q^2 +m_Z^2\cos 2\beta(I_q^{3L}
-e_q\sin^2\theta_W), \nonumber
\\&& M_{RR}^2 =m_{\tilde{U},\tilde{D}}^2 +m_q^2 +m_Z^2
\cos 2\beta e_q\sin^2\theta_W, \nonumber
\\&& M_{LR} =M_{RL} =\left(\begin{array}{ll} A_t -\mu\cot\beta &
(\tilde{q} =\tilde{t}) \\ A_b -\mu\tan\beta & (\tilde{q}
=\tilde{b}) \end{array} \right),
\end{eqnarray}
we used $m_{\tilde t_1}$, $A_t=A_b,\tan\beta$ and $\mu$ as the
input parameters. To simplify the calculations we assumed
$M_{\tilde{Q}}=M_{\tilde{U}} =M_{\tilde{D}}$, $m_{{\tilde
d}_{1,2}}=m_{{\tilde s}_{1,2}}=m_{{\tilde b}_{1}}+300$GeV, and
$m_{{\tilde u}_{1,2}}=m_{{\tilde c}_{1,2}}=m_{{\tilde
t}_{1}}+300$GeV. Such assuming of the relation between the squark
masses is done merely for simplicity, and actually, our numerical
results are not sensitive to the squark masses of the first and
second generation.
%, as shown below.

(ii)According to the experimental upper bound on the couplings in
the R-parity violating interaction\cite{c21}, we take the relevant
%\mbox{$\rlap{\kern0.3em/}R$ }
R-parity violating parameters as
$|\lambda^{''}_{132}|=1.00,|\lambda^{''}_{313}|=0.0026,|\lambda^{''}_{323}|=0.96$,
$|\lambda^{'}_{132}|=0.5,\ |\lambda^{'}_{323}|=0.9$.

In the numerical calculation we find that the contributions
induced by $\lambda'$ couplings are rather small, and they only
reach ${\cal O}(10^{-4})$. So we only discuss the effects induced
by $\lambda''$ below. In most of the previous studies, the bottom
quark mass was neglected, and $\Gamma_+=0$. In this paper we keep
the bottom quark mass, and one has $\Gamma_+\neq 0$. But the
results for transverse-plus rate $\delta\Gamma_+/\Gamma^0_+$ are
very small, which are around ${\cal O}(10^{-6})$, so we do not
show them in the curves. The other results are shown in Fig.5-13.

Fig.5 shows the dependence of the total corrections
$\delta\Gamma/\Gamma_0$ on $m_{\tilde{t}_1}$, assuming
$\tan\beta=40$, $\mu=300,500$ and $600$GeV, respectively. One
finds that the total corrections $\delta\Gamma/\Gamma_0$ increase
with decreasing $m_{\tilde{t}_1}$, and increase with increasing
$\mu$. They can reach about $4.0\%$, when $m_{\tilde{t}_1}=150$GeV
and $\mu=600$GeV. Comparing with the results in MSSM with
R-conservation reported in Ref.\cite{c15}, these corrections are
rather large and remarkable.

Fig.6 gives the total corrections $\delta\Gamma/\Gamma_0$ as a
function of $\mu$ for $m_{\tilde{t}_1}=150$GeV, $\tan\beta=4$ and
$40$, respectively. It should be noted that the magnitude of $\mu$
below $200$GeV have been ruled out by LEP II\cite{c22}. As shown
in this figure, we will not consider these areas. We find that the
corrections increase with increasing $\mu$  when $\mu>0$, and the
two curves have opposite trends when $\mu<0$.

In Fig.7 we show the total corrections $\delta\Gamma/\Gamma_0$ as
a function of $\tan\beta$ for $m_{\tilde{t}_1}=150$GeV, $\mu=-600$
and $200$GeV, respectively. We find that these corrections are not
sensitive to $\tan\beta$ for the smaller valus of $|\mu|$
($|\mu|=200$GeV), and sensitive to $\tan\beta$ for the larger
values of $|\mu|$ ($|\mu|=600$GeV), these features also can be
seen from Fig.6.

Fig.8 presents the dependence of the longitudinal corrections
$\delta\Gamma_L/\Gamma^0_L$ on $m_{\tilde{t}_1}$, assuming
$\mu=300$GeV, $\tan\beta=4$ and $40$, respectively. We find that
these corrections increase with decreasing $m_{\tilde{t}_1}$ , and
the longitudinal corrections are rather large, which can reach
about $3\%$. From this figure one can see that the main
contributions to the corrections to $\delta\Gamma/\Gamma_0$ come
from the longitudinal corrections.

Fig.9 shows the longitudinal corrections
$\delta\Gamma_L/\Gamma^0_L$ as a function of $\tan\beta$, for $
m_{\tilde{t}_1}=150$GeV, $\mu=-600$ and $200$GeV, respectively.
From this figure, one can see that the corrections are not
sensitive to $\tan\beta$. Comparing with Fig.6, these corrections
are plus when $\mu>0$ and minus when $\mu<0$.

Fig.10 presents the dependence of the longitudinal corrections
$\delta\Gamma_L/\Gamma^0_L$ on $\mu$, assuming $
m_{\tilde{t}_1}=150$GeV, $\tan\beta=4$ and $40$, respectively.
Note that the areas of $|\mu|<200$GeV has been ruled out by LEP
II. We can see that the magnitudes of these corrections increase
with increasing $|\mu|$.

The remainder of the figures show the transverse-plus
$\delta\Gamma_-/\Gamma^0_-$ as the functions of
$m_{\tilde{t}_1},\mu$ and $\tan\beta$, respectively. From
Fig.11-13 we can see that the magnitudes of the corrections
arising from the $\lambda''$ couplings are smaller, which only got
$0.4\%$. However, it is still important for the precise
measurements of the helicity content of the W gauge boson in top
quark decays in the future experiments.

To summarize, in general the corrections induced by $\lambda'$
couplings are negligibly small, but the corrections induced by
$\lambda''$ couplings may be important. Especially, the total
corrections $\delta\Gamma/\Gamma_0$ induced by the R-parity
violation couplings $\lambda''$ can reach about $4\%$, which are
smaller than the ${\cal O}(\alpha_s)$ radiative
corrections\cite{c13}, but are comparable with the corresponding
results in MSSM with R-parity conservation\cite{c15}. In addition,
The longitudinal corrections can reach about $3\%$, which are
still smaller than the results of the ${\cal O}(\alpha_s)$ QCD
corrections\cite{c13}, but larger than both of the ${\cal
O}(\alpha)$ EW and finite width corrections\cite{c14} and the SUSY
corrections with R-parity conservation\cite{c15}. The magnitudes
of the corrections to the transverse-minus are about $0.3\%$,
which are comparable with the results of \cite{c15} with R-parity
conservations, but are smaller than both of the ${\cal
O}(\alpha_s)$ QCD corrections and the ${\cal O}(\alpha)$ EW and
finite width corrections. The corrections to the transverse-minus
and the transverse-plus are too small to be observed, while the
total and longitudinal corrections induced by $\lambda''$
couplings may be observable in the precise measurements in top
quark decays in the future experiments; at least, interesting new
constraints on the R-parity violation couplings can be established.\\
\section*{Acknowledgments}

This work is supported in part by the National Natural Science
Foundation of China and the Specialized Research Fund for the
Doctoral Program of Higher Education.\vspace{.5cm}
\newpage
{\Large Appendix: \ \ \ Helicity Amplitudes}\\
(1)The explicit expressions of the helicity amplitudes of $A^v_1$
and $A^v_2$ in the case of $\lambda_b=-1$ are
\begin{eqnarray}
& &A^v_1(+,-,0)=\frac{ig}{8\sqrt{2}\pi^2}\frac{X}{M_W}\sqrt{(E_b-|\vec{p}_b|)E_t}(|\vec{p}_b|-E_W)\sin\frac{\theta}{2}C_0,\nonumber\\
& &A^v_1(+,-,-1)=\frac{-ig}{8\pi^2}X\sqrt{(E_b-|\vec{p}_b|)E_t}\cos\frac{\theta}{2}C_0,\nonumber\\
& &A^v_1(+,-,+1)=0,\nonumber\\
& &A^v_1(-,-,0)=\frac{-ig}{8\sqrt{2}\pi^2}\frac{X}{M_W}\sqrt{(E_b-|\vec{p}_b|)E_t}(|\vec{p}_b|-E_W)\cos\frac{\theta}{2}C_0,\nonumber\\
& &A^v_1(-,-,-1)=\frac{-ig}{8\pi^2}X\sqrt{(E_b-|\vec{p}_b|)E_t}\sin\frac{\theta}{2}C_0,\nonumber\\
& &A^v_1(-,-,+1)=0,\nonumber\\&
&A_2^v(+,-,0)=\frac{ig}{8\sqrt{2}\pi^2}Y\sqrt{(E_b-|\vec{p}_b|)E_t}\sin\frac{\theta}{2}\times\nonumber\\
& &\big\{-\frac{m_t^2}{M_W}|\vec{p}_b|(C_{12}+C_{22})
+\frac{E_W-|\vec{p}_b|}{M_W}C_{24}+\frac{m_t}{M_W}(E_W-|\vec{p}_b|)|\vec{p}_b|(C_{22}-C_{23})\big\},\nonumber\\
& &A_2^v(+,-,-1)=\frac{-ig}{8\pi^2}\sqrt{(E_b-|\vec{p}_b|)E_t}\cos\frac{\theta}{2}C_{24},\nonumber\\
& &A_2^v(+,-,+1)=0,\nonumber\\
& &A_2^v(-,-,0)=\frac{-ig}{8\sqrt{2}\pi^2}Y\sqrt{(E_b-|\vec{p}_b|)E_t}\cos\frac{\theta}{2}\times\nonumber\\
&
&\big\{-\frac{m_t^2}{M_W}|\vec{p}_b|(C_{12}+C_{22})+\frac{E_W-|\vec{p}_b|}{M_W}C_{24}+\frac{m_t}{M_W}|\vec{p}_b|(E_W-|\vec{p}_b|)(C_{22}-C_{23})\big\},\nonumber\\
& &A_2^v(+,-,-1)=\frac{-ig}{8\pi^2}\sqrt{(E_b-|\vec{p}_b|)E_t}\sin\frac{\theta}{2}C_{24},\nonumber\\
& &A_2^v(-,-,+1)=0.\nonumber
\end{eqnarray}
(2)The explicit expressions of the helicity amplitudes of $A^v_1$
and $A^v_2$ in the case of $\lambda_b=+1$ are
\begin{eqnarray}
A_1^v(+,+,0)&=&\frac{ig}{8\sqrt{2}\pi^2}X\sqrt{(E_b+|\vec{p}_b|)E_t}\frac{E_W+|\vec{p}_b|}{M_W}\cos\frac{\theta}{2}C_0,\nonumber\\
A_1^v(+,+,-1)&=&0,\nonumber\\
A_1^v(+,+,+1)&=&\frac{-ig}{8\pi^2}X\sqrt{(E_b+|\vec{p}_b|)E_t}\sin\frac{\theta}{2}C_0,\nonumber\\
A_1^v(-,+,0)&=&\frac{-ig}{8\sqrt{2}\pi^2}X\sqrt{(E_b+|\vec{p}_b|)E_t}\frac{E_W+|\vec{p}_b|}{M_W}\sin\frac{\theta}{2}C_0,\nonumber\\
A_1^v(-,+,-1)&=&0,\nonumber\\
A_1^v(-,+,+1)&=&\frac{-ig}{8\pi^2}X\sqrt{(E_b+|\vec{p}_b|)E_t}\cos\frac{\theta}{2}C_0,\nonumber
\end{eqnarray}
\begin{eqnarray}
A_2^v(+,+,0)&=&\frac{ig}{8\sqrt{2}\pi^2}Y\sqrt{(E_b+|\vec{p}_b|)E_t}\cos\frac{\theta}{2}\times\nonumber\\
\big\{-\frac{m_t^2}{M_W^2}|\vec{p}_b|(C_{12}+C_{22})&+&\frac{E_W+|\vec{p}_b|}{M_W}C_{24}+\frac{m_t}{M_W}|\vec{p}_b|(E_W+|\vec{p}_b|)(C_{22}-C_{23})\big\},\nonumber\\
A_2^v(+,+,-1)&=&0,\nonumber\\
A_2^v(+,+,+1)&=&\frac{-ig}{8\pi^2}Y\sqrt{(E_b+|\vec{p}_b|)E_t}\sin\frac{\theta}{2}C_{24},\nonumber\\
A_2^v(-,+,0)&=&\frac{-ig}{8\sqrt{2}\pi^2}Y\sqrt{(E_b+|\vec{p}_b|)E_t}\sin\frac{\theta}{2}\times\nonumber\\
\big\{-\frac{m_t^2}{M_W^2}|\vec{p}_b|(C_{12}+C_{22})&+&\frac{E_W+|\vec{p}_b|}{M_W}C_{24}+\frac{m_t}{M_W}|\vec{p}_b|(E_W+|\vec{p}_b|)(C_{22}-C_{23})\big\},\nonumber\\
A_2^v(-,+,-1)&=&0,\nonumber\\
A_2^v(-,+,+1)&=&\frac{-ig}{8\pi^2}Y\sqrt{(E_b+|\vec{p}_b|)E_t}\cos\frac{\theta}{2}C_{24}.\nonumber
\end{eqnarray}

\newpage
\begin{figure}[hbt]
\begin{center}
\epsfig{file=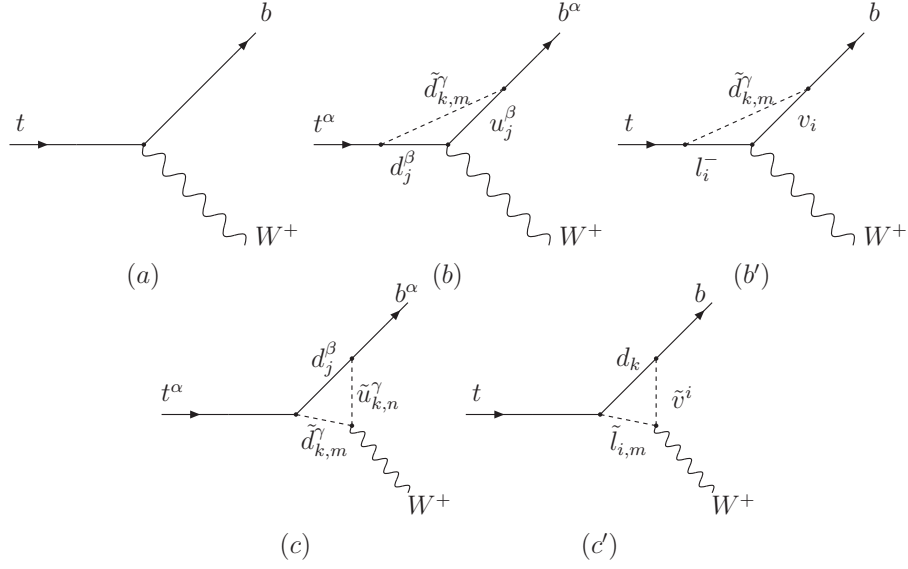,width=12cm}
 \caption{The Feynman diagrams including tree-level and vertex corrections
 for $t \rightarrow bW^+$. (b)(($b'$)) and (c)(($c'$)) are, respectively, for $\lambda''(\lambda')$.}
\end{center}
\end{figure}
\hspace{0.3cm}
\begin{figure}[hbt]
\begin{center}
\epsfig{file=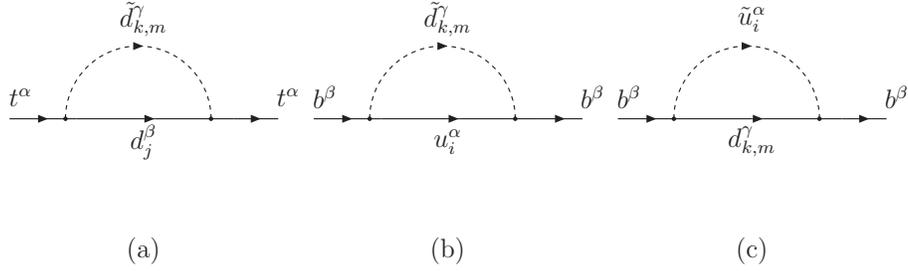,width=12cm}
 \caption{The self-energy diagrams of the top-quark and the bottom quark for $\lambda''$. (a) is for the top quark,
 (b) and (c) are for the bottom quark. }
\end{center}
\end{figure}
\hspace{0.3cm}
\begin{figure}[hbt]
\begin{center}
\epsfig{file=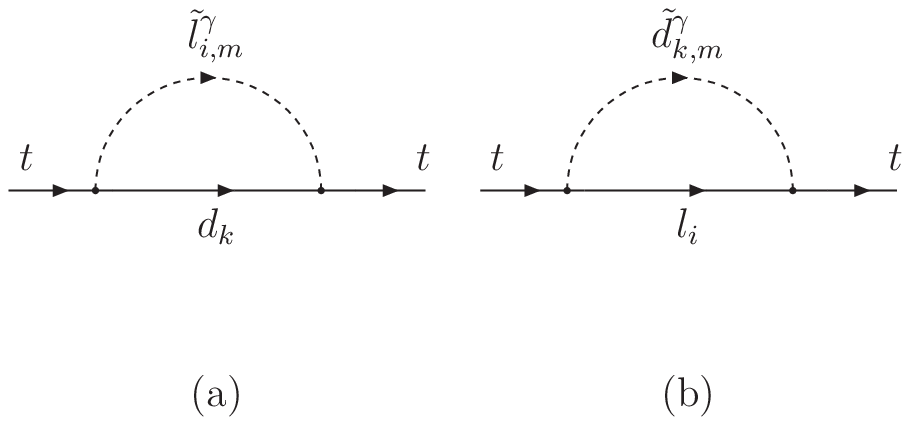,width=8cm}
 \caption{The self-energy diagrams of the top quark for $\lambda'$.}
\end{center}
\end{figure}
\hspace{0.3cm}
\begin{figure}[hbt]
\begin{center}
\epsfig{file=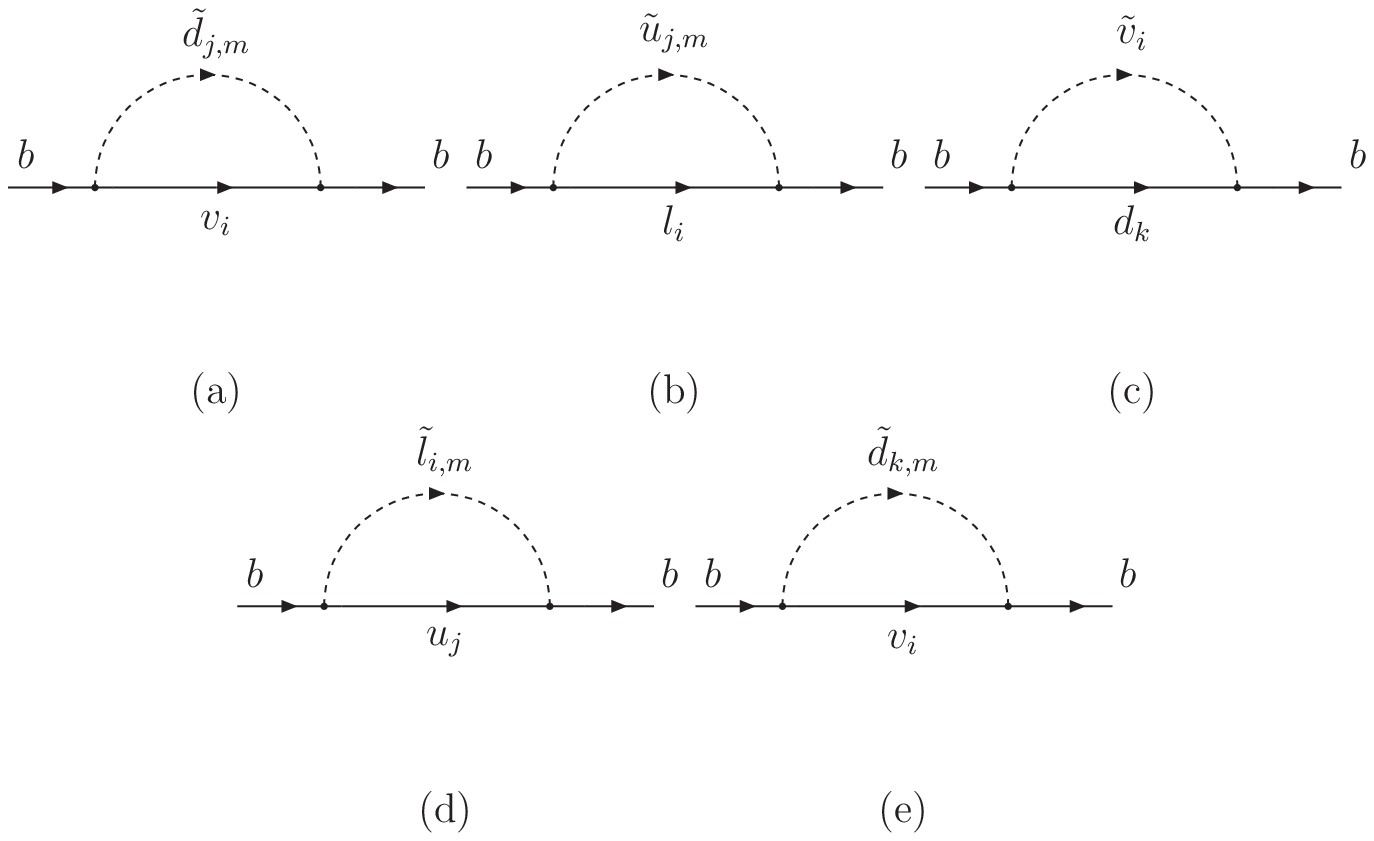,width=12cm}
 \caption{The self-energy diagrams of the bottom quark for $\lambda'$.}
\end{center}
\end{figure}
\hspace{0.3cm}
\begin{figure}[hbt]
\begin{center}
\epsfig{file=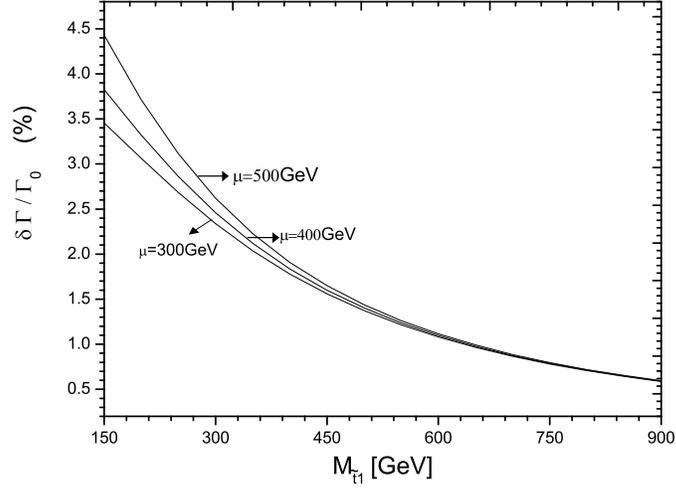,width=10cm} \caption{Dependence of the total
corrections $\delta\Gamma/\Gamma_0$ on the parameter
$m_{\tilde{t}_1}$, assuming $\tan\beta=40$, $A_t=A_b=1000$GeV and
$\mu=300,400,500$GeV, respectively}
\end{center}
\end{figure}
\begin{figure}[hbt]
\begin{center}
\epsfig{file=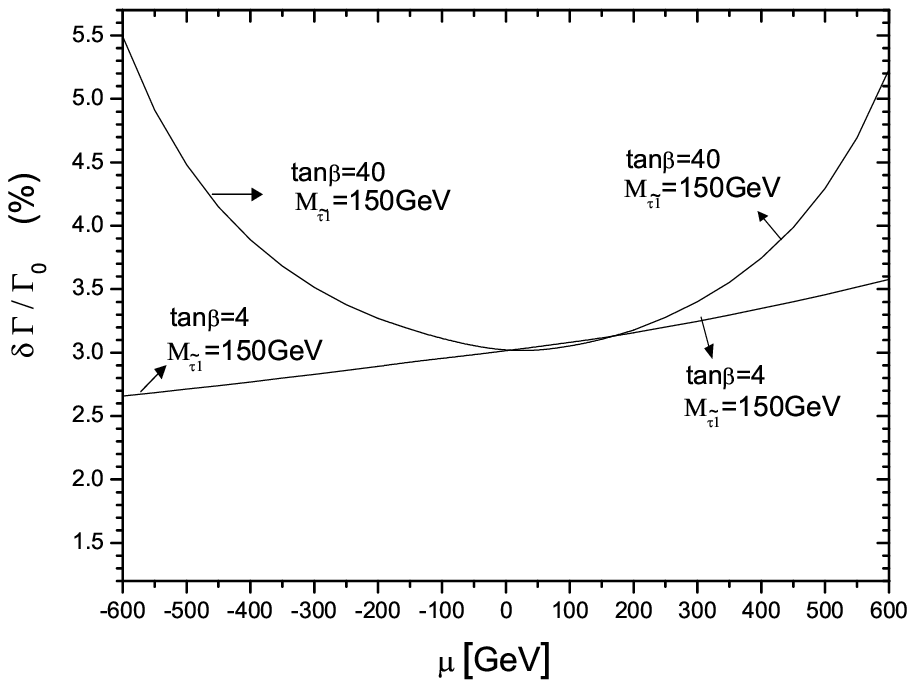,width=10cm}
 \caption{Dependence of the total
corrections $\delta\Gamma/\Gamma_0$ on the parameter $\mu$,
assuming $m_{\tilde{t}_1}=150$GeV, $A_t=A_b=1000$GeV and
$\tan\beta=4, 40$, respectively}
\end{center}
\end{figure}
\begin{figure}[hbt]
\begin{center}
\epsfig{file=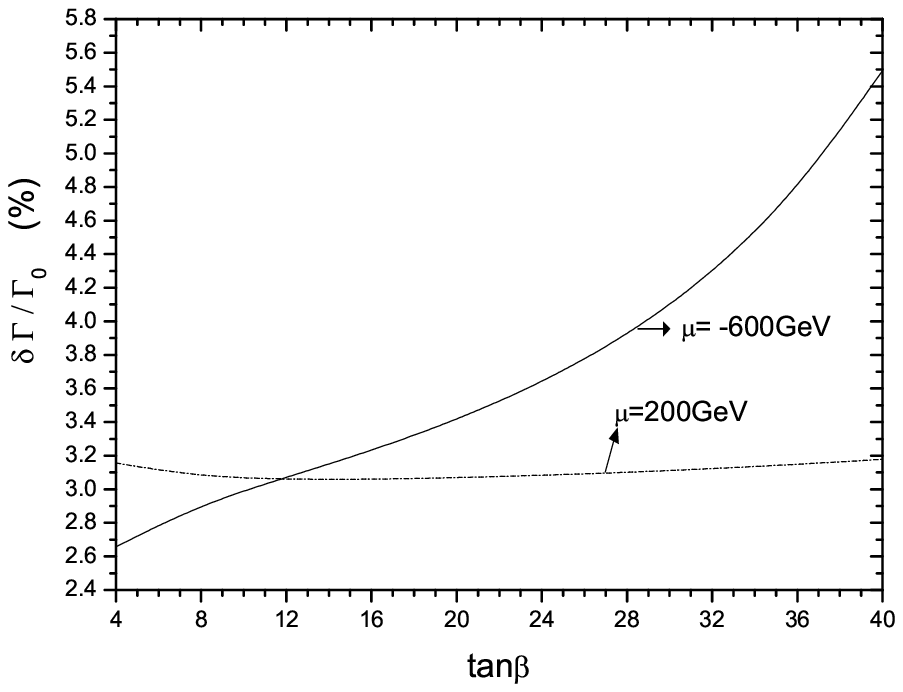,width=10cm} \caption{Dependence of the total
corrections $\delta\Gamma/\Gamma_0$ on the parameter $\tan\beta$,
assuming $m_{\tilde{t}_1}=150$GeV, $A_t=A_b=1000$GeV and
$\mu=-600, 200$GeV, respectively }
\end{center}
\end{figure}
\begin{figure}[hbt]
\begin{center}
\epsfig{file=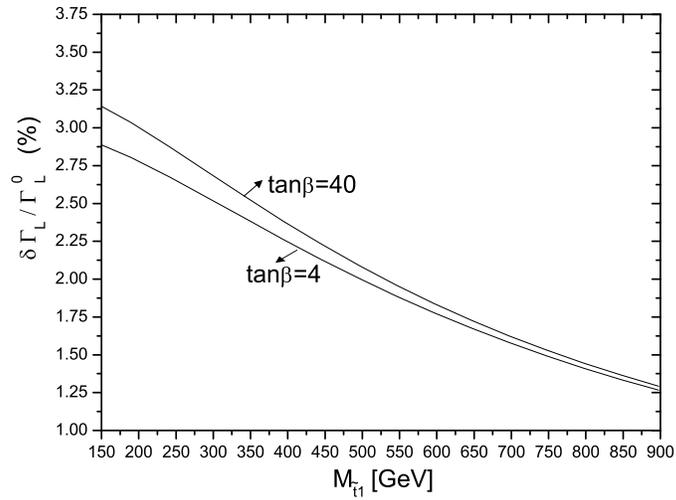,width=10cm} \caption{Dependence of the
longitudinal corrections $\delta\Gamma_L/\Gamma_L^0$ on the
parameter $m_{\tilde{t}_1}$, assuming $\mu=300$ GeV,
$A_t=A_b=1000$GeV and $\tan\beta=4,40$, respectively}
\end{center}
\end{figure}
\begin{figure}[hbt]
\begin{center}
\epsfig{file=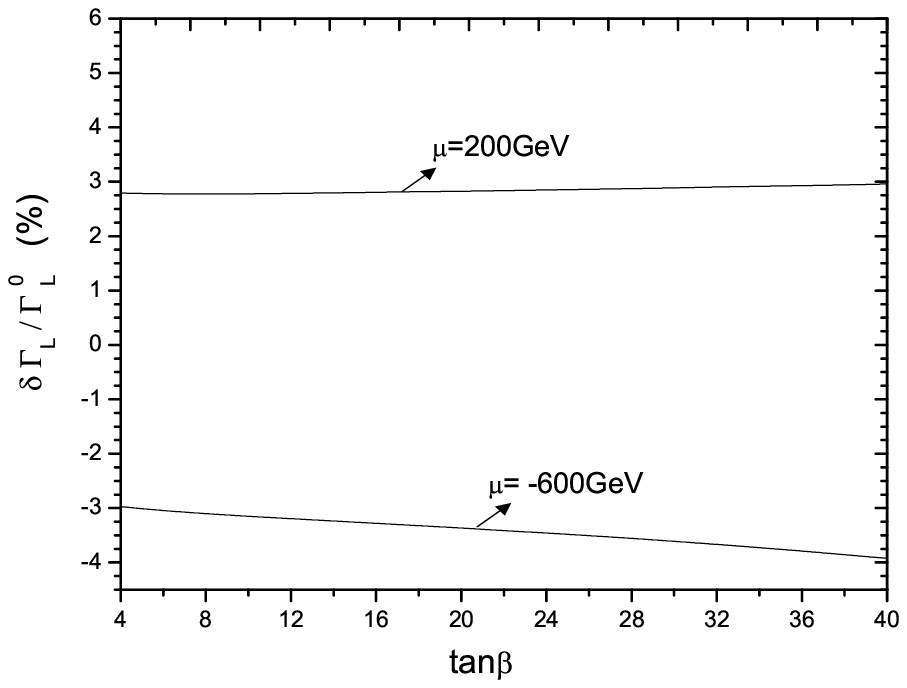,width=10cm} \caption{Dependence of the
longitudinal corrections $\delta\Gamma_L/\Gamma_L^0$ on the
parameter $\tan\beta$, assuming $m_{\tilde{t}_1}=150$GeV,
$A_t=A_b=1000$GeV and $\mu=-600, 200$GeV, respectively}
\end{center}
\end{figure}
\begin{figure}[hbt]
\begin{center}
\epsfig{file=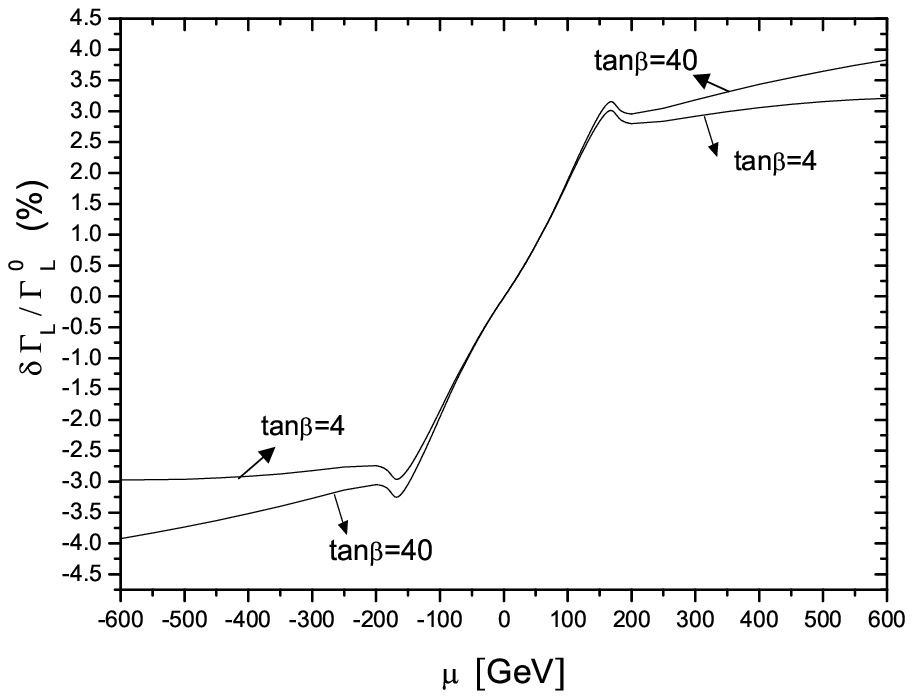,width=10cm} \caption{Dependence of the
longitudinal corrections $\delta\Gamma_L/\Gamma_L^0$ on the
parameter $\mu$, assuming $m_{\tilde{t}_1}=150$GeV,
$A_t=A_b=1000$GeV and $\tan\beta=4,40$. }
\end{center}
\end{figure}
\begin{figure}[hbt]
\begin{center}
\epsfig{file=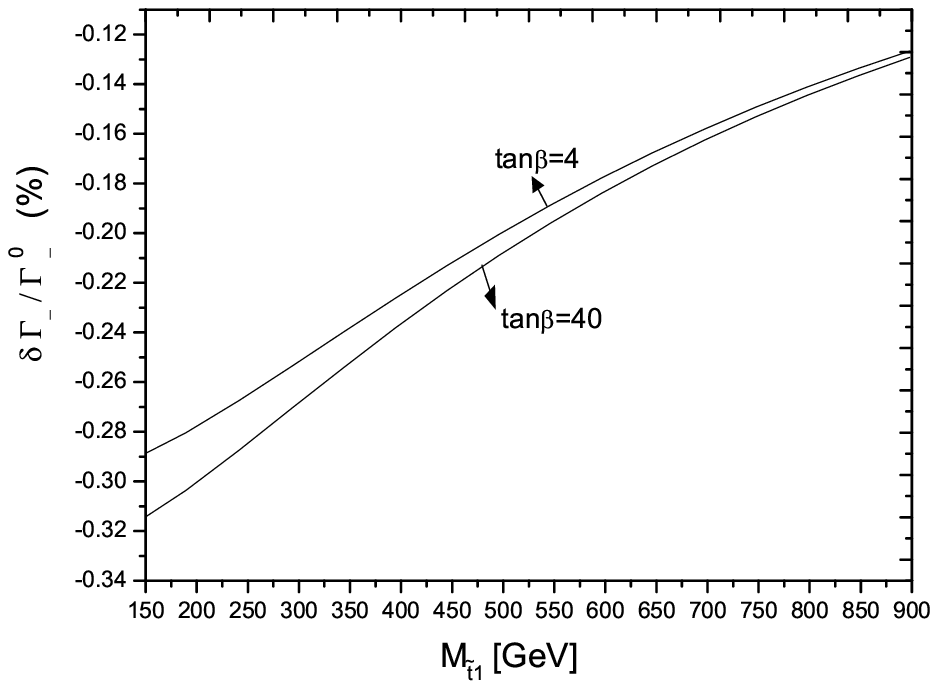,width=10cm} \caption{Dependence of the
transverse-minus corrections $\delta\Gamma_-/\Gamma_-^0$ on the
parameter $m_{\tilde{t}_1}$, assuming $\mu=300$GeV,
$A_t=A_b=1000$GeV and $\tan\beta=4,40$, respectively }
\end{center}
\end{figure}
\begin{figure}[hbt]
\begin{center}
\epsfig{file=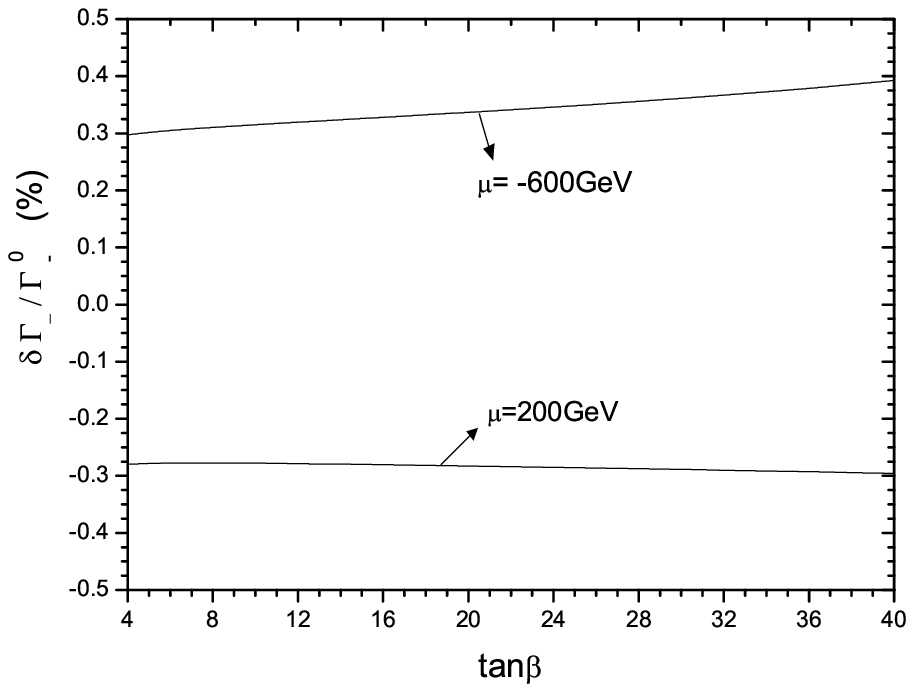,width=10cm} \caption{Dependence of the
transverse-minus corrections $\delta\Gamma_-/\Gamma_-^0$ on the
parameter $\tan\beta$, assuming $m_{\tilde{t}_1}=150$GeV,
$A_t=A_b=1000$GeV and $\mu=-600, 200$GeV, respectively }
\end{center}
\end{figure}
\begin{figure}[hbt]
\begin{center}
\epsfig{file=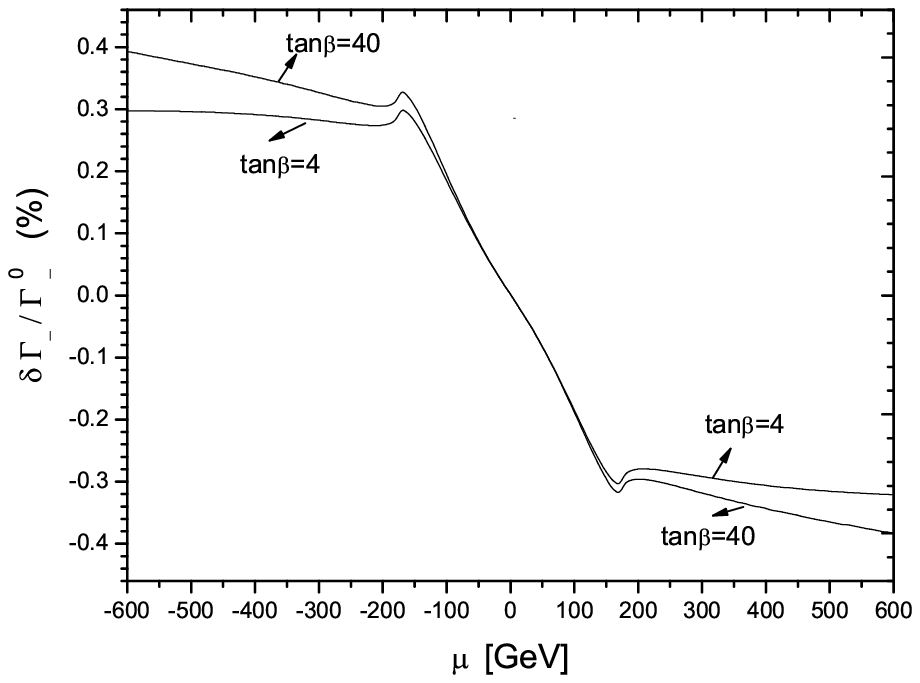,width=10cm} \caption{Dependence of the
transverse-minus corrections $\delta\Gamma_-/\Gamma_-^0$ on the
parameter $\mu$, assuming $m_{\tilde{t}_1}=150$GeV,
$A_t=A_b=1000$GeV and $\tan\beta=4, 40$, respectively}
\end{center}
\end{figure}

\end{document}